\documentclass[a4paper]{article}

\usepackage{INTERSPEECH2020}
\usepackage{subfigure}

\title{Dual-Path Transformer Network: Direct Context-Aware Modeling for End-to-End Monaural Speech Separation}
\name{Jingjing Chen$^1$, Qirong Mao$^{1,2}$, Dong Liu$^1$}
\address{
  $^1$School of Computer Science and Communication Engineering, Jiangsu University, China\\
  $^2$Jiangsu Engineering Research Center of big data ubiquitous perception and intelligent agriculture applications, Zhenjiang, China}
\email{2221808071@stmail.ujs.edu.cn, mao\_qr@ujs.edu.cn, 2111908002@stmail.ujs.edu.cn}

\begin{document}

\maketitle
\begin{abstract}
  The dominant speech separation models are based on complex recurrent or convolution neural network that model speech sequences indirectly conditioning on context, such as passing information through many intermediate states in recurrent neural network, leading to suboptimal separation performance. In this paper, we propose a dual-path transformer network (DPTNet) for end-to-end speech separation, which introduces direct context-awareness in the modeling for speech sequences. By introduces a improved transformer, elements in speech sequences can interact directly, which enables DPTNet can model for the speech sequences with direct context-awareness. The improved transformer in our approach learns the order information of the speech sequences without positional encodings by incorporating a recurrent neural network into the original transformer. In addition, the structure of dual paths makes our model efficient for extremely long speech sequence modeling. Extensive experiments on benchmark datasets show that our approach outperforms the current state-of-the-arts (20.6 dB SDR on the public WSj0-2mix data corpus). 
\end{abstract}
\noindent\textbf{Index Terms}: direct context-aware modeling, transformer, dual-path network, speech separation, deep learning

\section{Introduction}

Speech separation, often referred to as the cocktail party problem \cite{bronkhorst2000cocktail,haykin2005the}, is a fundamental task in signal processing with a wide range of real-world applications, such as separating clean speech from noisy speech signals to improve the accuracy of automatic speech recognition. The human auditory system has the remarkable ability to extract separate sources from a complex mixture, while this task seems to be difficult for automatic calculation system, especially when only a monaural recording of mixed-speech is available.

Although there are many challenges in
monaural speech separation, a lot of attempts have been made in previous works to deal with this problem over the decades. Before the deep learning era, many traditional methods are introduced for this task, such as non-negative matrix factorization (NMF) \cite{schmidt2006single-channel,le2015deep}, computational auditory scene analysis (CASA) \cite{wang2008computational} and probabilistic models~\cite{virtanen2006speech}. However, these models usually only work for closed-set speakers, which significantly restricts their practical applications. With the success of deep learning techniques on various domains \cite{gou2018sparsity,ocquaye2019dual},
researches start to design data-based models to separate the mixture of unknown speakers, which overcomes the obstacles of the traditional methods. In general, deep learning techniques for monaural speech separation can be divided into two categories: time-frequency (T-F) domain methods and end-to-end time-domain approaches. Based on T-F features created by calculating the short-time Fourier transform (STFT), T-F methods separate the T-F features for each source and then reconstruct the source waveforms by inverse STFT \cite{hershey2016deep,chen2017deep,yu2017permutation,kolbaek2017multitalker,yang2019improved}. They usually use the original phase of mixture to synthesize the estimated source waveforms, which retain the phase of the noisy mixture. This strategy imposes an upper limit on the separation performance. To overcome this problem, time-domain approach is proposed in paper \cite{luo2018tasnet} , which directly model the mixture waveform using an encode-decoder framework and has made great progress in recent years \cite{luo2019conv,shi2019deep,shi2019end,takahashi2019recursive,ditter2019multi,luo2019dual,zeghidour2020wavesplit}.

\begin{figure*}[t]
  \centering
  \includegraphics[width=\linewidth]{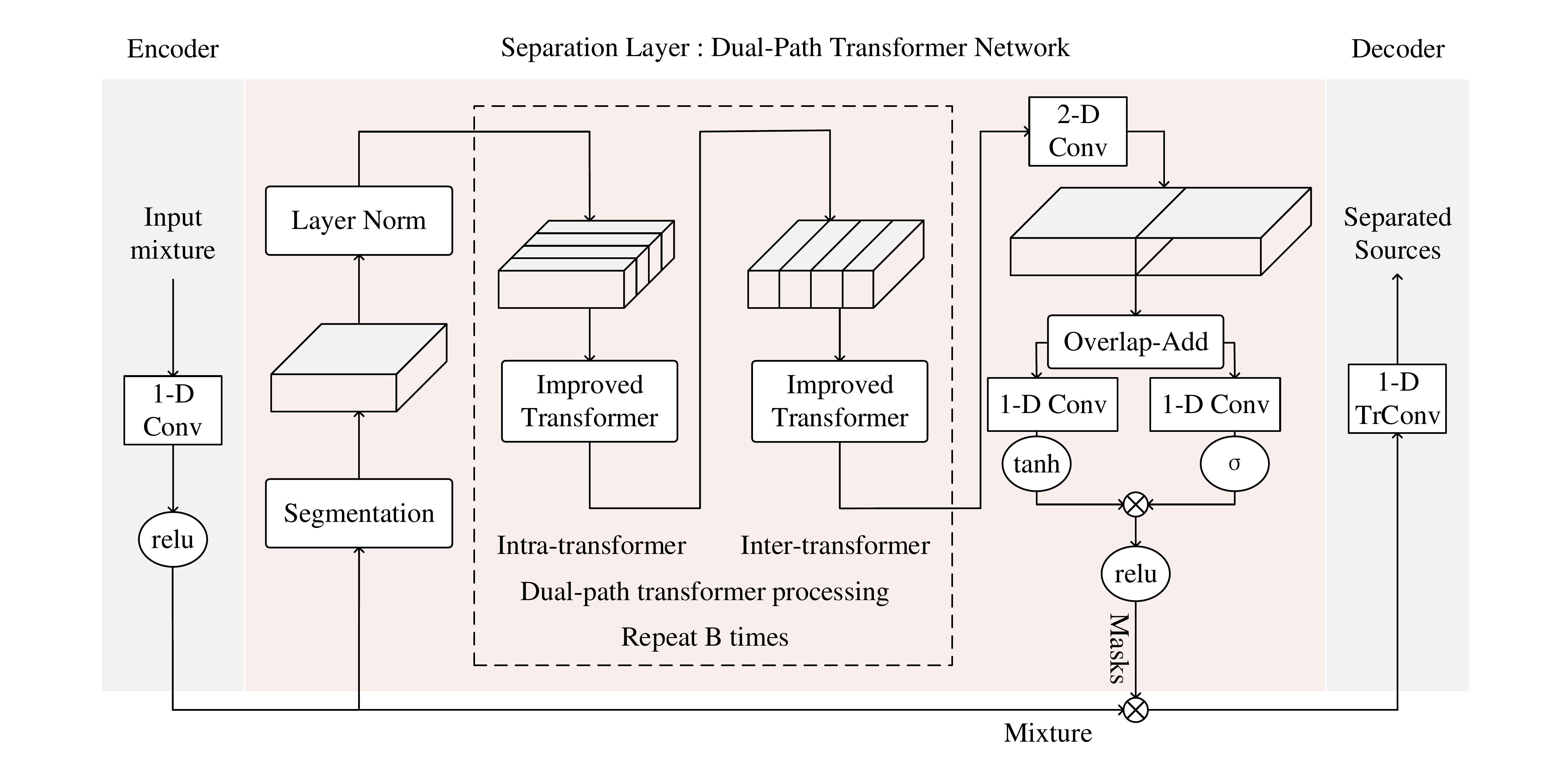}
  \caption{Framework of speech separation with dual-path transformer network.}
  \label{fig:dpt}
\end{figure*}

However, the dominant speech separation models are usually based on recurrent neural network (RNN) or convolution neural network (CNN), which cannot model the speech sequences directly conditioning on context~\cite{sperber2018self}, leading to suboptimal separation performance. For example, RNN based models need to pass information through many intermediate states. And the models based CNN suffer from the problem of limited receptive fields. Fortunately, the transformer based on self-attention mechanism can resolve this problem effectively~\cite{vaswani2017attention}, in which elements of the inputs can interact directly. Nevertheless, the transformer usually only deals with sequences with length of hundreds, while end-to-end time-domain speech separation systems often model extremely long input sequences, which can sometimes be tens of thousands. Dual-path network is an effective method to deal with this problem~\cite{luo2019dual}.

Inspired by the above, we propose a dual-path transformer network (DPTNet) for end-to-end monaural speech separation, which introduces a improved transformer to allow direct context-aware modeling on the speech sequences, leading to superior separation performance. The major contributions of our work are summarized as follows.

\begin{itemize}
\item[1.] To the best of our knowledge, this is the first work that introduces direct context-aware modeling into speech separation. This method enables the elements in speech sequences can interact directly, which is beneficial to information transmission.
\item[2.] We integrate a recurrent neural network into original transformer to make it can learn the order information of the speech sequences without positional encodings. And we embed this improved transformer into a dual-path network, which makes our approach efficient for extremely long speech sequence modeling.
\item[3.] Extensive experiments on benchmark datasets show that our approach outperforms the current state-of-the-arts (20.6 dB SDR on the public WSj0-2mix data corpus).
\end{itemize}


The remains of this paper are organized as follows. We introduces monaural speech separation with DPTNet in Section 2, present the experiment procedures in Section 3, analyze the experiment results in Section 4, conclude this paper and indicate future work in Section 5.

\section{Speech separation with dual-path transformer network}

As depicted in Figure~\ref{fig:dpt}, our speech separation system consists of three stages: encoder, separation layer and decoder, which is similar to that of Conv-TasNet in paper~\cite{luo2019conv}. First, an encoder is used to convert segments of the mixture waveform into corresponding features in an intermediate feature space. Then the features are feed to the separation layer to construct a mask for each source. Finally, the decoder reconstructs the source waveforms by converting the masked features. In the following, we outline the encoder and decoder, and describe the separation layer, namely our dual-path transformer network, in detail.

\subsection{Encoder}

If we denote the speech mixture by $x\in R^{1\times T}$, then we can divide it into overlapping vectors $\bm{x}\in R^{L\times I}$ of length $L$ samples, where $I$ is the number of vectors. The encoder receive $\bm{x}$ and output the speech signal $X\in R^{N\times I}$ as follows:
\begin{equation}
  X=ReLU(\bm{x}*W)
  \label{eq_encoder}
\end{equation}
where the encoder can be characterized as a filter-bank $W$ with $N$ filters of length $L$, which is actually a 1-D convolution module.

\subsection{Separation layer: dual-path transformer network}

The separation layer, namely dual-path transformer network, is composed of three stages: segmentation, dual-path transformer processing and overlap-add, which is inspired by the common dual-path network~\cite{luo2019dual}.

\subsubsection{Segmentation}
Firstly, the segmentation stage splits $X$ into overlapped chunks of length $K$ and hop size $H$. Then all the chunks are concatenated to be a 3-D tensor $D\in R^{N\times K\times P}$.

\subsubsection{Dual-path transformer processing}

Broadly speaking,
the transformer is composed of an encoder and a decoder~\cite{vaswani2017attention}. The encoder and decoder share the same model structure, except that the decoder is a left-context-only version for generation. To avoid confusions, the transformer in this paper refers specially to the encoder part, and it is comprised of three core modules: scaled dot-product attention, multi-head attention and position-wise feed-forward network.

\begin{figure}[t]
  \centering
  \includegraphics[width=\linewidth]{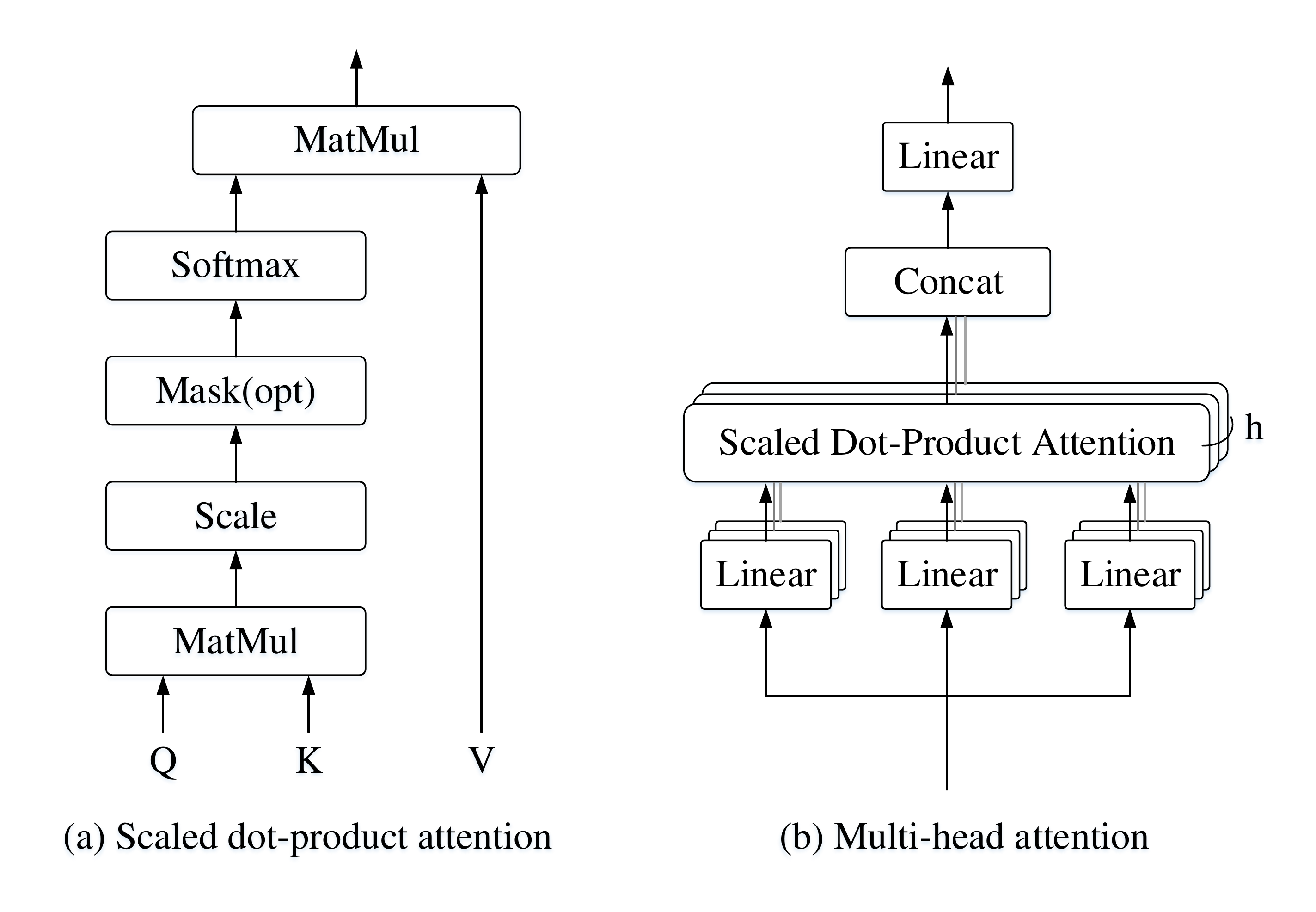}
  \caption{Attention mechanism of the transformer.}
  \label{fig:attention}
\end{figure}

\begin{figure}[t]
  \centering
  \includegraphics[width=\linewidth]{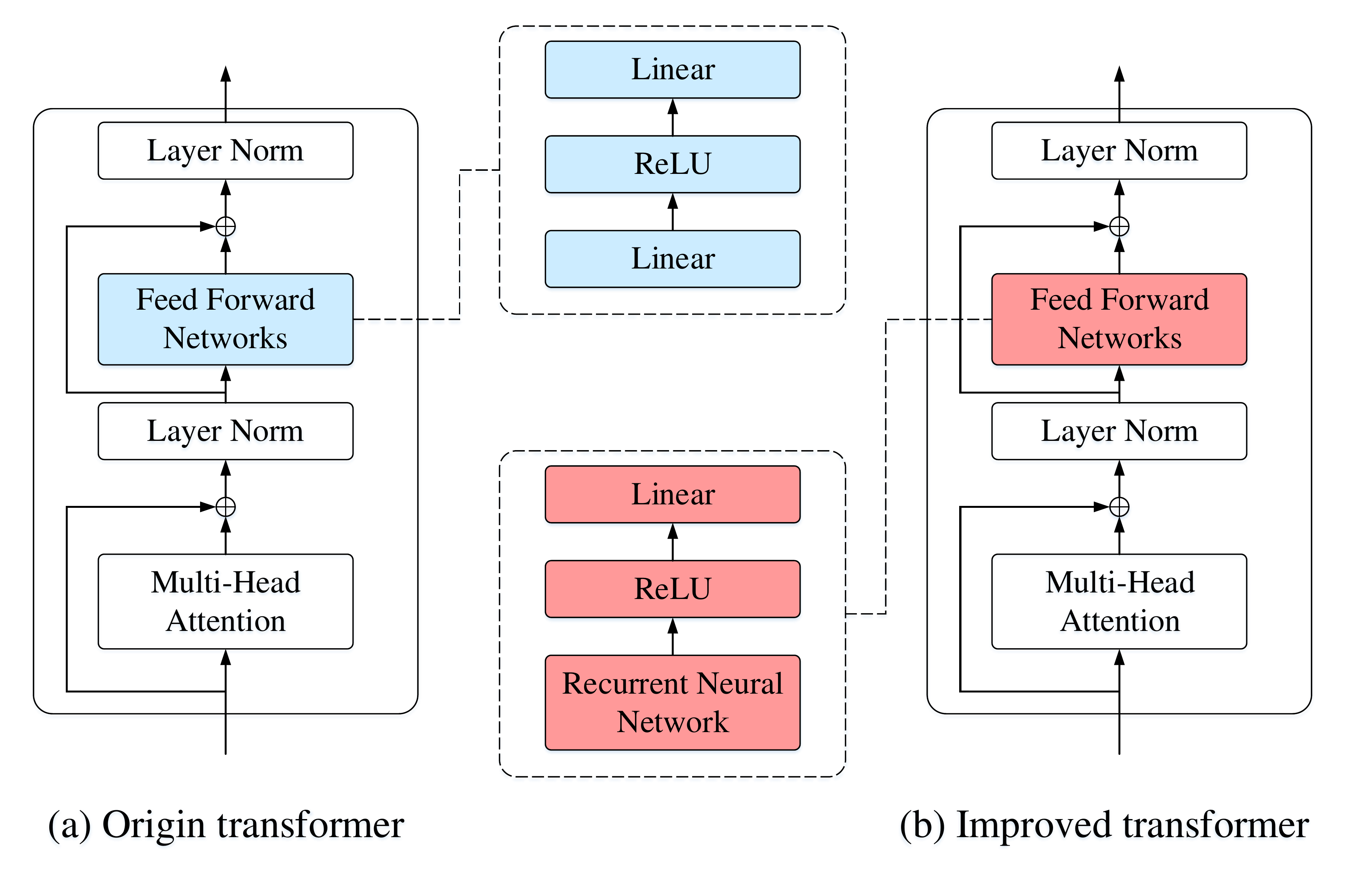}
  \caption{Architecture of the origin and improved transformers.}
  \label{fig:transformers}
\end{figure}

Scaled dot-product attention is an effective self-attention mechanism that associate different positions of input sequences to calculate representations for the inputs, which is shown in Figure~\ref{fig:attention}(a). The final output of this module is computed as a weighted sum of the values, where the weight for each value is computed by a attention function of the query with the corresponding keys. Multi-head attention is composed of multiple scaled dot-product attention modules, as depicted in Figure~\ref{fig:attention}(b). First, it linearly maps the inputs $h$ times with different, learnable linear projections to get parallel queries, keys and values respectively. Then the scaled dot-product attention is performed on these mapped queries, keys and values simultaneously. Position-wise feed-forward network is a fully connected feed-forward network. It is comprised of two linear transformations with a $ReLU$ activation in between. Besides the three core modules, transformer also includes several residual and normalization layers. We present the overall structure of the transformer in Figure~\ref{fig:transformers}(a) and it can be formulated as follows:
\begin{equation}
  Q_{i}=ZW^{Q}_{i}, K_{i}=ZW^{K}_{i}, V_{i}=ZW^{V}_{i} \quad i\in [1,h]
  \label{equ:equ_1}
\end{equation}
\begin{equation}
  \begin{split}
  head_{i}&=Attention({Q}_{i},{K}_{i},{V}_{i})\\
  &=softmax(\frac{Q_{i}K^{T}_{i}}{\sqrt{d}})V_{i}
  \label{equ:equ_2}
  \end{split}
\end{equation}
\begin{equation}
  MultiHead\!=\!Concat(head_{1}\!,\!...,\!head_{h})W^{O}
  \label{equ:equ_3}
\end{equation}
\begin{equation}
  Mid=LayerNorm(Z+MultiHead)
  \label{equ:equ_4}
\end{equation}
\begin{equation}
  FFN=ReLU(MidW_{1}+b_{1})W_{2}+b_{2}
  \label{equ:equ_5}
\end{equation}
\begin{equation}
  Output=LayerNorm(Mid+FFN)
  \label{equ:equ_6}
\end{equation}
Here, $Z\in{R^{l\times d}}$ is the input with length $l$ and dimension $d$, and $Q_{i}, K_{i}, V_{i} \in{R^{l\times d/h}}$ are the mapped queries, keys and values. $W^{Q}_{i}, W^{K}_{i}, W^{V}_{i}\in{R^{d\times d/h}}$ and $W^{O}\in{R^{d\times d}}$ are parameter matrices.
$FFN$ denotes the output of the position-wise feed-forward network, in which $W_{1}\in R^{d\times d_{ff}}$, $W_{2}\in R^{d_{ff}\times d}$, $b_{1}\in R^{d_{ff}}$, $b_{2}\in R^{d}$, and $d_{ff}=4\times d$.

The elements in speech sequences modeled by the transformer can contact directly without intermediate transmission, which introduces direct context-aware modeling into our method.

One thing missed by the above transformer is how to utilize the order information in the speech sequences. The origin transformer adds positional encodings to the input embeddings to represent order information, which is sine-and-cosine functions or learned parameters. However, we find that the positional encodings are not suitable for dual-path network and usually lead to model divergence in the training process. To learn the order information, we replace the first fully connected layer with a recurrent neural network in the feed-forward network, which is a interesting improvement from paper~\cite{sperber2018self}:
\begin{equation}
  FFN=ReLU(RNN(Mid))W_{2}+b_{2}
  \label{equ:dqu_7}
\end{equation}
We show this improved transformer in Figure~\ref{fig:transformers}(b) and apply it in next dual-path transformer processing stage.

In dual-path transformer processing stage, the output $D$ of the segmentation stage is passed to a heap of $B$ dual-path transformers (DPTs), as presented in Figure~\ref{fig:dpt}. Each DPT consists of intra-transformer and inter-transformer, which are committed to modeling local and global information respectively. The intra-transformer processing block first model the local chunk independently, which acts on the second dimension of $D$:
\begin{equation}
\begin{split}
  D{^{intra}_b}&=IntraTransformer_b[D{^{inter}_{b-1}}]\\
  &=[transformer(D{^{inter}_{b-1}}[:,:,i]),i=1,...,P]
\end{split}
  \label{eq_dpt_1}
\end{equation}
Then the inter-transformer is used to summarize the information from all chunks to learn global dependency with performing on the last dimension of $D$:
\begin{equation}
\begin{split}
  D{^{inter}_b}&=InterTransformer_b[D{^{intra}_b}]\\
  &=[transformer(D{^{intra}_b}[:,j,:]),j=1,...,K]
\end{split}
  \label{eq_dpt_2}
\end{equation}
where $b=1,...,B$ and $D{^{inter}_0}=D$. Note that the layer normalization in each sub-transformer is applied to all dimensions.

Indeed, this structure makes each element in speech sequences interact directly with only some elements and interact with the rest elements through an intermediate element. This fact imposes a slight negative impact on the direct context-aware modeling. However, the structure of dual paths allows our approach to model for extremely long speech sequences efficiently. In general, the small shortcoming caused by the dual-path structure is far less than the benefits it brings.

\subsubsection{Overlap-Add}
The output of the last inter-transformer $D{^{inter}_B}$
is used to learn a mask for each source by a 2-D convolution layer. The masks are transformed back into sequences $M_s\in R^{N\times I}$ by overlap-add, and masked encoder features for $s$-th source are obtained by the element-wise multiplication between $X$ and $M_s$:
\begin{equation}
  Y_s=X\cdot M_s
  \label{eq_multi}
\end{equation}

\subsection{Decoder}

In decoder, a transposed convolution module is used to reconstruct separated speech signals $\bm{y}_s\in R^{L\times I}$ for $s$-th source:
\begin{equation}
  \bm{y}_s=Y_s*V
  \label{eq_decoder}
\end{equation}
where values in $V\in R^{N\times L}$ are the parameters of the transposed convolution module. Then the overlap-add method is applied to obtain the final waveforms $y_s\in R^{1\times T}$. The structure and function of decoder are both symmetrical with those of the encoder.

\section{Experiment}

\subsection{Dataset}

We evaluate our proposed model on two-speaker speech separation using the WSJ0-2mix~\cite{hershey2016deep} and LS-2mix dataset~\cite{panayotov2015librispeech}.

The WSJ0-2mix dataset is derived from the WSJ0 data corpus~\cite{garofolo1993csr}. The 30 hours of training data and 10 hours of validation data contain two-speaker mixtures generated by randomly selecting utterances from different speakers in the WSJ0 training set si\_tr\_s, and mixing them at random signal-to-noise ratios (SNR) between -5 dB and 5 dB. 5-hours test set is similarly generated using utterances from unseen speakers in WSJ0 validation set si\_dt\_05 and evaluation set si\_et\_05.

LS-2mix is created based on the Librispeech dataset~\cite{panayotov2015librispeech}, which is a new corpus of reading English speech.
Two speakers are randomly selected from the train-100 set to generate training mixtures, at various SNRs uniformly sampled between 0 dB and 5 dB. The validation and test set are similarly generated using utterances from unseen speakers in the Librispeech validation and test set. Generated LS-2mix dataset contains 20000, 5000 and 3000 utterances in the train/validation/test set.

\subsection{Experiment setup}

In encoder and decoder, the window size is 2 samples and a 50\% stride size is used. The number of filters is set to 64. As for the separation layer, the number of dual-path transformers, namely $B$, is set to 6, and $h=4$ parallel attention layers are employed.

In the training stage, we train proposed model for 100 epochs on 4-second long segments, and the criteria for early stopping is no decrease in the loss function on validation set for 10 epochs. Adam \cite{kingma2014adam} is used as the optimizer and gradient clipping with maximum L2-norm of 5 is applied during training. We increase the learning rate linearly for the first $warmup\_n$ training steps, and then decay it by 0.98 for every two epochs:
\begin{equation}
\begin{split}
  lrate&=k_{1}\cdot d^{-0.5}_{model}\cdot n\cdot warmup\_n^{-1.5}\\
  &\quad\quad\quad\quad\quad\quad\quad\quad\quad\quad when\ n\le warmup\_n\\
  &=k_{2}\cdot 0.98^{epoch//2}\quad\quad when\ n>warmup\_n\\
\end{split}
  \label{eq_lr}
\end{equation}
where $n$ is the step number, $k_{1}$, $k_{2}$ are tunable scalars, and $k_{1}=0.2$, $k_{2}=4e^{-4}$, $warmup\_n=4000$ in this paper.

These hyper-parameters are selected empirically according to the setups in the dual-path network~\cite{luo2019dual} and transformer~\cite{vaswani2017attention}. A Pytorch implementation of our DPTNet model can be found at ``https://github.com/ujscjj/DPTNet".

\subsection{Training objective}

We train proposed model with utterance-level permutation invariant training (uPIT) \cite{kolbaek2017multitalker} to maximize scale-invariant source-to-noise ratio (SI-SNR) \cite{luo2018tasnet}. SI-SNR is defined as:

\begin{equation}
  s_{target}=\frac{\langle\tilde x, x\rangle x}{\lVert{x}\rVert^2}
  \label{eq_si-snr_1}
\end{equation}
\begin{equation}
  e_{noise}=\tilde x-s_{target}
  \label{eq_si-snr_2}
\end{equation}
\begin{equation}
  SI-SNR:=10log_{10}\frac{\lVert{s_{target}}\rVert^2}{\lVert{e_{noise}}\rVert^2}
  \label{eq_si-snr_3}
\end{equation}
where $x$, $\tilde x$ are clean and estimated source respectively, both of which are normalized to zero-mean before the calculation.

\section{Performance evaluation}


\begin{table}[t]
  \caption{Comparison with other methods on WSJ0-2mix in SI-SNR (dB), SDR (dB) and Model Size}
  \label{tab:results_1}
  \centering
  \begin{tabular}{lccc}
    \toprule
    \textbf{Method}      & \textbf{SI-SNR}      & \textbf{SDR}      & \textbf{Model Size}\\
    \midrule
    DPCL++ \cite{isik2016single}                   & 10.8            & -               & 13.6M\\
    uPIT-BLSTM-ST \cite{kolbaek2017multitalker}    & -               & 10.0            & 92.7M\\
    Deep Attractor \cite{chen2017deep}             & 10.5            & -               & -\\
    ADANet \cite{luo2018speaker}                   & 10.4            & 10.8            & 9.1M\\
    Grid LSTM PIT \cite{xu2018single}              & -               & 10.2            & -\\
    ConvLSTM-GAT \cite{li2018cbldnn}               & -               & 11.0            & -\\
    Chimera++ \cite{wang2018alternative}           & 11.5            & 12.0            & -\\
    WA-MISI-5 \cite{wang2018end}                   & 12.6            & 13.1            & 32.9M\\
    BLSTM-TasNet \cite{luo2018tasnet}              & 13.2            & 13.6            & -\\
    Conv-TasNet-gLN \cite{luo2019conv}             & 15.3            & 15.6            & 5.1M\\
    Conv-TasNet+MBT \cite{lam2019mixup}            & 15.5            & 15.9            & -\\
    Deep CASA \cite{liu2019divide}                 & 17.7            & 18.0            & 12.8M\\
    FurcaNeXt \cite{zhang2020furcanext}            & -               & 18.4            & 51.4M\\
    DPRNN \cite{luo2019dual}                       & 18.8            & 19.0            & \textbf{2.6M}\\
    \midrule
    \textbf{DPTNet}                                & \textbf{20.2}   & \textbf{20.6}   & \textbf{2.69M}\\
    \bottomrule
  \end{tabular}
\end{table}

\begin{table}[t]
  \caption{Comparison with baselines on the LS-2mix dataset}
  \label{tab:results_2}
  \centering
  \begin{tabular}{lccc}
    \toprule
    \textbf{Method}      & \textbf{SI-SNR}      & \textbf{SDR}      & \textbf{Model Size}\\
    \midrule
    Conv-TasNet-gLN \cite{luo2019conv}             & 12.9            & 13.5            & 5.1M\\
    DPRNN \cite{luo2019dual}                       & 15.0            & 15.6            & \textbf{2.6M}\\
    \midrule
    \textbf{DPTNet}                                & \textbf{16.2}   & \textbf{16.8}   & \textbf{2.69M}\\
    \bottomrule
  \end{tabular}
\end{table}

%
%

In all experiments, we report the scale-invariant signal-to-noise (SI-SNR) and signal-to-distortion ratio (SDR) to measure the separation performance of our DPTNet, both of which are often employed in various speech separation systems.

We first report the SI-SNR and SDR scores on the WSJ0-2mix dataset obtained by our model and the well-known separation methods in recent years.
As shown in Table~\ref{tab:results_1}, our DPTNet achieves 20.2 dB and 20.6 dB on the metrics of SI-SNR and SDR respectively, where a new state-of-the-art performance is achieved. Benefiting from the direct context-aware modeling, the elements in the speech sequences modeled by our DPTNet can interact directly, which results in the optimal monaural speech separation performance. In addition, our model maintains a small model size.

To prove the generalization of our approach, we conduct related experiments on the LS-2mix dataset. Compared to those in the WSJ0-2mix data corpus, the mixtures in LS-2mix is difficult to separate, but this does not interfere with the comparison between our method and the baselines. We reproduce two classical methods, namely Conv-TasNet~\cite{luo2019conv} and DPRNN~\cite{luo2019dual}, as baselines. Table~\ref{tab:results_2} lists the average SI-SNR and SDR obtained by our DPTNet and the two baselines, where our direct context-aware modeling is still significantly superior to the state-of-the-art approach DPRNN. This presents the generalization of our method and further demonstrates the effectiveness of it.

\section{Conclusion and future work}

In this paper, we propose the dual-path transformer network for end-to-end multi-speaker monaural speech separation, which models the speech sequences directly conditioning on context. Our model can learn the order information in speech sequences without positional encodings and model effectively for extremely long sequences of speech signals. Experiments on two benchmark datasets demonstrate the effectiveness of proposed model, and we achieve a new state-of-the-art performance on the public WSJ0-2mix data corpus. In the future, we would like to extend this work by directly modeling long speech feature sequences without the dual-path structure. It is promising to further improve the separation performance.

\bibliographystyle{IEEEtran}

\bibliography{mybib}


\end{document}